\documentclass{epl}

\usepackage{graphicx}
\usepackage{graphics}
\usepackage{amsmath}
\usepackage{amssymb}
\usepackage{hyperref}
\usepackage{latexsym}

\def\be{\begin{equation}}
\def\ee{\end{equation}}
\def\bea{\begin{eqnarray}}
\def\eea{\end{eqnarray}}

\def\mx{\mbox{\boldmath $x$\unboldmath}}
\def\my{\mbox{\boldmath $y$\unboldmath}}

\def\Q{\mbox{\boldmath $Q$\unboldmath}}
\def\I{\mbox{\boldmath $I$\unboldmath}}
\def\nabbold{\mbox{\boldmath $\nabla$\unboldmath}}

\def\sigbold{\mbox{\boldmath $\sigma$\unboldmath}}

\def\sigac{\mbox{\boldmath $\sigma$\unboldmath$^a$}}

\def\sigd{\mbox{\boldmath $\sigma$\unboldmath$^d$}}
\def\sigr{\mbox{\boldmath $\sigma$\unboldmath$^{\tiny r}$}}

\title{Shear flow induced  isotropic to nematic transition in a suspension of active filaments}
\shorttitle{Shear-stabilised active filaments}
\author{Sudipto MUHURI \inst{1}\thanks{E-mail: \email{sudipto@rri.res.in}},
Madan RAO\inst{1,2}\thanks{E-mail: \email{madan@rri.res.in}}
\and
Sriram RAMASWAMY\inst{3}\thanks{E-mail: \email{sriram@physics.iisc.ernet.in}}}
\shortauthor{S. MUHURI \etal}
\institute{
  \inst{1} Raman Research Institute, Bangalore 560 080 India\\ 
  \inst{2} National Centre for Biological Sciences, Tata Institute of Fundamental Research, 
  Bangalore 560 065 India\\ 
  \inst{3} Centre for Condensed Matter Theory, Department of Physics,
Indian Institute of Science, Bangalore 560012, India
}

\begin{document}

\maketitle

\begin{abstract}
We study the effects of externally applied shear flow on a model of suspensions 
of motors and filaments, via the equations of active hydrodynamics  
[PRL {\bf 89} (2002) 058101; {\bf 92} (2004) 118101]. In the absence of shear, the orientationally ordered phase of {\it both} polar and apolar active particles is always unstable 
at zero-wavenumber. An imposed steady shear large 
enough to overcome the active stresses stabilises 
both apolar and moving polar phases. Our work is relevant to {\it in vitro} studies of active filaments, 
the reorientation of endothelial cells
subject to shear flow and  shear-induced motility of attached cells.
\end{abstract}

The collective behaviour of suspensions of active particles\footnote{By 
active particle we mean an object which absorbs energy from its 
surroundings and dissipates it in the process of carrying out internal movements; 
examples include self-propelled organisms, motor-filament complexes, and 
agitated monolayers of granular particles.} 
is fundamentally different from that of their passive counterparts, 
as highlighted by recent studies on hydrodynamic instabilities 
\cite{aditisriram,tanni}, rheology \cite{hat} and 
steady state patterns \cite{kardarlee,kruse,gautam} of active particle systems. 
Recent experimental studies on model systems such as motor-filament 
extracts \cite{kaes,asters} and 
suspensions of swimming bacteria \cite{albert,shiva,goldstein} 
or spermatocytes \cite{howard} have also reported 
strong departures 
from {\it equilibrium} behaviour. 

One surprising prediction \cite{aditisriram} is that, unlike in passive
suspensions, long-range uniaxial orientational order in 
active Stokesian\footnote{where viscosity dominates and inertia 
is ignored} suspensions of polar particles is {\em always} destroyed by a  
hydrodynamic instability. This is because the flow generated by a 
small, long-wavelength splay or bend deformation imposed on an oriented 
configuration always acts instantaneously (in the Stokesian regime) 
to further increase the deformation
\cite{aditisriram,rheo-orient}.

If only this instability could be controlled, 
we would have an orientationally ordered phase of active
particles with rather unusual rheological properties
\cite{rheo-orient}. We already noted two dramatic rheological consequences
\cite{hat,tanni}: (i) the orientationally ordered  phase is characterised by a
nonzero steady-state average of the deviatoric stress -- a kind of yield stress --
and (ii) on approaching the orientationally ordered state from the isotropic
fluid, a suspension of active {\it contractile} elements exhibits solid-like
behaviour {\it without translational arrest}.  However to obtain this novel
material we would first need to stabilise this orientationally ordered phase.

In this paper we ask whether an imposed shear flow can suppress this
instability of active Stokesian suspensions. We address this question by
constructing a set of coarse-grained equations for the hydrodynamics of the
active particle suspensions. We consider apolar (nematic) as well as polar
order, keeping in mind that the latter correspond to phases with a nonzero
drift velocity of the active particles.  Here is a summary of our results,
valid in both $2$ and $3$ dimensions.  (i) The instability obtained in
\cite{aditisriram} occurs for {\em apolar} ordered suspensions as well; (ii)
Imposed shear larger than a critical strain rate ${\dot{\gamma}}_c$ stabilises
the orientationally ordered phase yielding a stability diagram (Fig.\,4)
controlled by two variables, a flow alignment parameter $\lambda$ and the  the
ratio of the shear to active stress.  
 The drift velocity of the 
vector-ordered phase depends on ${\dot{\gamma}}$ and $\lambda$.  
Possible experimental tests and
some complications regarding boundary conditions and back-flow are discussed at
the end of the paper. We now turn to the details of our calculations.

{\it (A) Hydrodynamic equations} : The hydrodynamic variables for a suspension
of active particles \cite{aditisriram,hat} are the total momentum density ${\bf
g} = \rho {\bf u}$ of the particles $+$ fluid, where ${\bf u}$ is the
hydrodynamic velocity field and $\rho$ the density, the
concentration $c({\bf  r},t)$ of active particles, and the orientational 
order parameter field. For {\em polar} systems, the order parameter 
is the polarisation vector ${\bf p}({\bf r},t)$ of the force dipoles 
associated with the active particles. Active systems being out of 
thermal equilibrium, a local polarisation always implies a local  
average drift velocity of the active particle relative to the fluid. 
We thus take ${\bf p}$ to be the velocity field of the active particles 
relative to the fluid. 
{\em Apolar} nematic orientational order is characterised by a traceless symmetric 
tensor $\Q \equiv S ({\bf p} {\bf p} - p^2\,(1/d) \I)$, where ${\bf p}$ is now the director and $S$ is the scalar nematic order parameter \cite{degennes}.
We focus on the polar case, and indicate differences where 
appropriate for the apolar case.  
Momentum conservation in the Stokesian limit means
$\nabbold \cdot \sigbold = 0$, where the
total stress tensor $\sigbold \equiv \sigac + \sigr + \sigd$ is 
the sum of contributions from activity, order-parameter gradients and viscous dissipation, 
and overall incompressibility of the suspension means $\nabbold \cdot {\bf u} = 0$, 
$\rho = \rho_0$, the mean density.  

The equations for the active polar order parameter ${\bf p}$ 
read
\begin{eqnarray}
\label{order}
&\partial_t {\bf  p}& + \left( {\bf  u} \cdot {\nabbold} \right) {\bf  p} 
-\frac{1}{2} \left( \nabbold \times {\bf  u} \right) \times {\bf  p}
 + \left[
\lambda_{1} \left({\bf  p}\cdot \nabbold \right) {\bf  p} + \lambda_{2}
\left(\nabbold \cdot {\bf  p} \right) {\bf  p} + \lambda_{3} \nabbold \vert
{\bf  p}\vert^{2} \right] \nonumber\\&=&
\frac{\lambda}{2}\left({\nabbold}\,{\bf u}+ ({ \nabbold}\,{\bf u})^{T}\right)
\cdot {\bf  p}-\zeta\, \nabbold c({\bf r},t) + \Gamma\, {\bf
h}\,
 + \ldots
\end{eqnarray}

In (\ref{order}), the first three terms on the left are 
the material derivative 
of ${\bf p}$ (co-moving and co-rotating with the suspension) and 
the square brackets (as well as the $\zeta$ term on the right) 
contain symmetry-allowed 
{\em polar} contributions \cite{tonertu}
(of which the $\lambda_1$ term alone \cite{tanni} is active),
ruled out in apolar nematohydrodynamics, whether 
passive \cite{degennes,landau} or active \cite{aditisriram}. 
The first term on the right, together with the $\nabbold \times {\bf u}$ 
term on the left, lead to flow-alignment \cite{degennes}.
The relaxation dynamics is contained in the order-parameter
molecular field 
${\bf  h} = c(r,t) \left[ \alpha {\bf  p} - \beta \vert{\bf p}\vert^{2} {\bf  p} + K \nabla^{2} {\bf  p} \right]
$
which favours a fixed length for ${\bf p}$ and assigns an elastic cost to 
inhomogeneities in the one-Frank-constant approximation.

Conservation of active particles states
\begin{equation}
\label{conc}
\partial_t c=-  \nabbold \cdot \left[\,c\,({\bf  u} + {\bf  p})\right]. 
\end{equation}

The reactive stress from order-parameter gradients is
\begin{equation}
\label{sigmar}
\sigr= 
-\frac{\lambda}{2}\left({\bf p}\,{\bf h}+ ({\bf h}\,{\bf p})^T\right) + 
\Pi  \, {\bf I} 
\end{equation}
where $\Pi$ is a generalised pressure; 
and the viscous stress is 
$\sigd =\frac{\eta}{2} \,\left({\nabbold}\,{\bf u}+ ({\nabbold}\,{\bf u})^{T}\right) \equiv \eta \,{\bf A}$,
where $\eta$ the bare shear viscosity of the suspension.

To determine the active 
stress, we make use of the fact that
the simplest active particle, on long
timescales, is a permanent force dipole \cite{aditisriram,hat}.
To leading order the deviatoric part of the stress coming from activity is given by
\begin{equation}
\label{actstress}
{\sigac} \left({\bf  r}, t\right) = W c\left({\bf  r}, t\right)\left({\bf p} \,{\bf p} - 
p^2\frac{1}{d} {\bf I}\right)\, ,
\end{equation}
where $d$ is the spatial dimension, and the magnitude and {\it sign} of $W$ characterise the nature of the elementary force dipoles \cite{aditisriram,hat}.

Here we use this hydrodynamic description 
to study the effect of shear flow on the ordering and stability of active particle
suspensions. 
\begin{figure} 
\begin{center}
\includegraphics[width=3in,angle=-90]{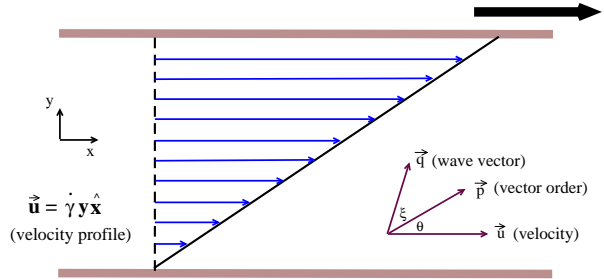}
\caption{Suspension of active vector ordered particles subject to an imposed
velocity along the ${\hat \mx}$-axis, $u_0={\dot \gamma} y {\hat \mx}$ with a
gradient along the ${\hat \my}$-axis. The shear alignment angle $\theta$ and
the angle $\xi$ between the wavevector ${\bf q}$ and 
the ordering direction are also
shown.}
 
\end{center}
 
\end{figure}

{\it (B) Shear flow} : We impose a planar shear flow (Fig.\,1) along the ${\hat
{\bf{\mx}}}$-axis, with a velocity gradient along ${\hat {\bf {\my}}}$, giving
rise to an imposed velocity field $ {\bf u}_0 = {\dot \gamma} y {\hat {\bf
{\mx}}}$. 

The instability of the orientationally ordered phase  \cite{aditisriram} and
its stabilisation due to shear is most simply seen in two dimensions
($d=2$); we therefore present detailed calculations in $d=2$, and merely state
results in $d=3$. 

We look for homogeneous steady states of (\ref{order}), (\ref{conc});
 the steady state concentration is a uniform $c(r,t) = c_0$. 
The amplitude and phase of the steady state active vector order parameter 
are, respectively,  
\begin{eqnarray}
p_{0}^2   &   =   &  \frac{\alpha}{\beta} + \frac{\dot{\gamma}}{2\beta}{\sqrt{\lambda^{2}-1 }} \\
\tan \theta  & =  & \sqrt{\frac{\lambda - 1}{\lambda + 1}}\, .
\label{steady} 
\end{eqnarray}

The flow alignment parameter $\lambda$ can take values between $1$ and
$\infty$, corresponding to $0 \leq \theta \leq \pi/4$. Note that 
while $p_0$ {\it increases} with shear rate, the phase $\theta$
is {\it independent} of it.

{\it (C) Stability of orientational order} : To determine the stability of this
homogeneous steady state, we set ${\bf p}({\bf r},t) = p_{0} (\cos \theta, \sin
\theta) + \delta {\bf p}({\bf r},t)$, ${\bf u}({\bf r},t) = {\bf u}_0 + \delta
{\bf u}({\bf r},t)$
 and $c({\bf r},t) = c_0 + \delta c({\bf r},t)$, where the
perturbations 
are assumed small. It is convenient to decompose $\delta {\bf p}$ and  $\delta
{\bf u}$ parallel and perpendicular to the ordering direction ${\bf p}_0$,
e.g., $\delta {\bf p} = (\delta p_{\parallel}, \delta {p}_{\perp})$, 
and to eliminate the ``massive'' field $\delta {p}_{\parallel}$ 
in favour of the remaining variables yielding, 
after a spatial fourier transform, 

\begin{eqnarray}
&&\left[\partial_{t} + iq_{\parallel}\lambda_{1}p_{0}+ 
2iq_{\perp}F\lambda_{3}p_{0}+{\dot{\gamma}}\sqrt{\lambda^{2}-1}\right]\delta p_{\perp}(q) 
+ \frac{\dot{\gamma}}{2}\left(F\cos\theta -\sin\theta\right)\frac{\partial}{\partial q_y}
\left(q_{\perp}\delta p_{\perp}(q)\right)\nonumber\\
&&=\left[\frac{ip_{0}q^{2}}{2\rho_{0} q_{\parallel}}
\left(1+\lambda\cos2\xi\right)\right]\delta g_{\perp}(q) + 
\left[\frac{i\dot{\gamma}q_{\parallel}}{2}D_{0} + 
\frac{i\lambda\dot{\gamma}q_{\parallel}}{2}D_{0}\cos2\theta\right]\delta c_q,  
\end{eqnarray}
where $P = 2\Gamma \alpha +\dot{\gamma}\left({\lambda}^{2}-1\right)$, 
$F={\dot{\gamma}}/P$ and $D_0={\zeta}/P$. 
Since the instability we are interested in has maximum growth rate \cite{aditisriram} 
at zero wavenumber, we ignore the term $q_i \frac{\partial}{\partial q_j}$ 
operating on the fields, whose effect can be shown to vanish 
for $q\rightarrow 0$.

Fluctuations in the hydrodynamic 
velocity field,  ${\delta {\bf u}} \equiv {\delta {\bf g}}/\rho_0$, 
to linear order are governed by the Stokes equation. Imposing 
incompressibility and eliminating the pressure, 
eliminating $\delta p_{||}$ in favour of the remaining fields as 
above, and fourier-transforming in space, we find  
$q_{\parallel}=q_{x}\cos\theta +q_{y}\sin\theta$ to obtain,
\begin{eqnarray}
0 = &-&\eta q^{2}\delta {g}_{\perp}(q) + \left(\frac{iWq_{\parallel}^2 {q}_{\perp}}{q^2}\right) \delta c_q+\left[\frac{2iWc_{0}q_{\parallel}{q}_{\perp}}{p_{0}q^{2}}\right] {q}_{\perp} \delta {p}_{\perp}(q)-\frac{iWc_{0}q_{\parallel}}{p_{0}}\delta {p}_{\perp}(q)\, .
\end{eqnarray}

Concentration fluctuations to linear order are given by
\begin{equation}
\partial_{t}\delta c_{q} = - i\,q_{\parallel}\,p_{0} \delta c_q -ic_{0}\left(q_{\perp}-Fq_{\parallel}\right)\delta p_{\perp}(q)\, .
\end{equation}

We are now in a position to compute the full fluctuation spectrum. While we have analysed the linear
stability of the orientationally ordered phase in the basis spanned by $\delta
{\bf u}$, $\delta {\bf p}$ and $\delta c$, we find that the origin of
instability and its recovery by the shear flow can be understood even in the
absence of the concentration equation. Thus, to make the subsequent analysis
more transparent, we drop the concentration fluctuation terms at the
outset. 

Following \cite{aditisriram}, we express the fluctuation spectrum in terms of
the splay fluctuation $\Phi=\nabla_{\perp}\delta p_{\perp}$, the in-plane
expansion rate $\Theta=\nabla_{\perp}\delta g_{\perp}$,
and the angle $\xi$ made by ${\bf  q}$ with the ordering direction:
\begin{equation}
\left[\partial_{t} + iq_{\parallel}\lambda_{1}p_{0}+{\dot{\gamma}}\sqrt{\lambda^{2}-1}\left(1-\frac{1}{4\lambda}\right) + O({\dot \gamma}^2)\right]\Phi_{q} -\left[\frac{ip_{0}q^{2}}{2\rho_0 q_{\parallel}}\left(1+\lambda\cos2\xi\right)\right]\Theta_{q} = 0
\end{equation}
and 
\begin{equation}
0 = -\eta q^{2}\Theta_{q}-\left[\frac{iWc_{0}q_{\parallel}\cos2\xi}{p_{0}}\right]\Phi_{q}
\end{equation}
This implies that the splay fluctuations $\Phi_{q}$ have a growth rate
\begin{equation}
\Omega = \frac{Wc_{0}}{2\eta}\cos2\xi\left(1 + \lambda\cos2\xi\right)-
\dot{\gamma}\sqrt{\lambda^{2}-1}\left(1-\frac{1}{4\lambda}\right) + O({\dot \gamma}^2) \, .
\label{dispersion}
\end{equation}
First recall the generic instability when ${\dot \gamma}=0$: 
the oriented phase is {\it always} unstable, either to
splay or to bend fluctuations, depending on the sign of $W$ \cite{aditisriram}.
This can be seen {\it even at $q=0$},
where the growth rate $\Omega_{+}(0) >0$ for $-\pi/4>\xi>\pi/4$ when $W>0$, and 
$\Omega_{-}(0) >0$ for $\pi/4 > \xi > 3\pi/4$, when $W<0$ (Fig.\ 2).
The dispersion curve $\Omega_{\pm}$ is shown in Fig.\,3 :
fluctuations with wavenumber $q<q_0 \propto \vert W \vert$  grow in time; this 
sets the scale over which orientational order can be stabilised. 
  Note that {\it both} the polar phase, as noted by \cite{aditisriram}, and the apolar nematic as well, are generically unstable in the absence of shear. This is clear, since the polar terms containing $\{\lambda_i\}$ and $\zeta$ in (\ref{order}), do not appear in the growth rate equation (\ref{dispersion}).
\begin{figure}
\begin{center}
\includegraphics[width=3in,angle=-90]{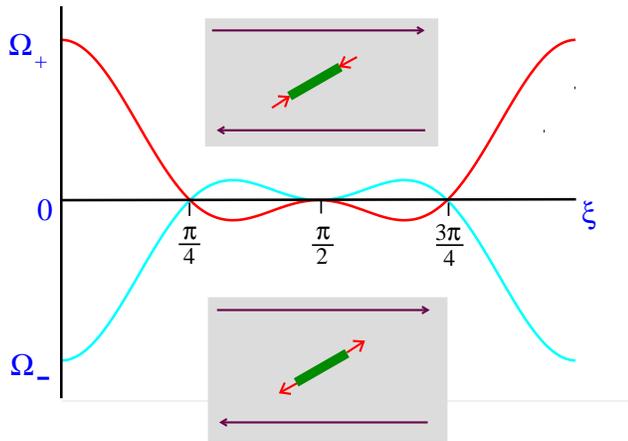}
\caption{The growth rates $\Omega_{\pm}(q=0)$ versus $2\xi$ 
(angle between perturbing and the ordering direction), 
for either sign of activity $W$. Inset shows the induced flow lines arising 
from active rods subject to a flow.}
\end{center}
\end{figure}

The imposed shear flow represented by a horizontal line 
${\dot \gamma} = \mbox{const.}$ in Fig.\,3, cuts this dispersion curve 
at $q^{*}({\dot \gamma})$, suggesting that fluctuations whose scale is 
{\it smaller} than $1/q^{*}$ are the first to be stabilised by the shear flow.
Our estimate of $q^*$ is qualitative :
the crossover from $\Omega$-dominated 
to $\dot{\gamma}$-dominated at nonzero $q$ cannot strictly be evaluated 
in our treatment, since we dropped the $q_x \partial{q_y}$ terms on 
the grounds that they wouldn't matter at $q = 0$ where the instability 
is fastest. 
As ${\dot \gamma}$ is increased, $q^{*}({\dot \gamma})$ decreases, till at
${\dot \gamma}={\dot \gamma}_c$, this cutoff scale 
moves to zero, as $q^* = ({\dot \gamma} - {\dot \gamma}_c)^{1/2}$.
At this shear rate ${\dot \gamma}_c$, the oriented phase is 
completely stabilised by the shear flow.
This defines a stability boundary as a function of $\lambda$ and $W$.

The stability phase diagram is best represented by defining a dimensionless 
{\it active} Peclet number, $Pe_a=2\eta \dot{\gamma}/{\vert W \vert c_{0}}$, as 
the ratio of the imposed shear rate to the typical shear-rate produced 
around the active particles. As one crosses from the unstable to the stable 
region in the plane of $Pe_a$ and flow-alignment parameter $\lambda$ (Fig.\,4) 
the orientational order parameter sets in at the value 
given by Eq. (\ref{steady}), which in effect gives a discontinuous transition 
since the order parameter in the hydrodynamically unstable region is zero.   
For polar active 
particles, the shear-stabilised oriented phase has a nonzero drift of the particles 
with respect to the solvent.

Note that the critical shear rate required to stabilise the oriented phase
${\dot \gamma}_c$ is larger for positive $W$ than negative. This is consistent
with our earlier observation \cite{hat}, that the flow induced
by the active
stresslets, oppose the
imposed flow when $W>0$ and enhance it when $W<0$.

\begin{figure}
\onefigure[height=5.1cm,angle=-90]{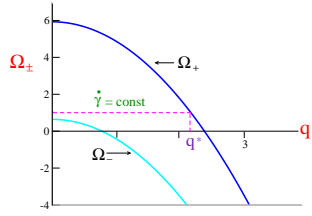}
\caption{Dispersion curves $\Omega_{\pm}$ versus $q$, for
$\xi$ corresponding to maximum growth rate. 
The horizontal line corresponding to a frequency
$\dot{\gamma}\sqrt{\lambda^{2}-1}\left(1-\frac{1}{4\lambda}\right)$
defines the scale 
$q^*$ beyond which fluctuations are stabilised. 
As ${\dot \gamma}$ increases towards ${\dot \gamma}_c$, the stabilisation 
scale goes to zero as $({\dot \gamma} - {\dot \gamma}_c)^{1/2}$.
}
\label{dispersion}
\end{figure}

\begin{figure}
\begin{center}
\includegraphics[width=3in,angle=-90]{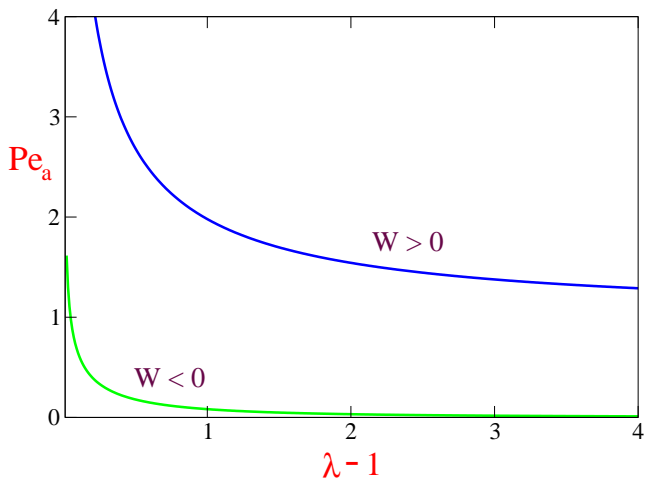}
\caption{Stability diagram in the plane of Active Peclet number $Pe_a$ and 
flow-alignment parameter $\lambda$.}
\end{center}
\end{figure}

{\it (D) Shear flow induced stabilisation in} $d=3$ : The calculation of the
stability diagram is more tedious in $d=3$; however since the spirit is the
same, we merely quote results and point out differences. To start with, we note
that ${\bf p}_0$, the steady state orientational conformation in the presence
of the shear flow ${\bf u}_0 = {\dot \gamma} y {\hat {\bf {\mx}}}$ is the same
as (\ref{steady}), i.e.. it still lies in the plane of the imposed velocity and
its gradient (the $xy$ plane).  We then decompose the fluctuations in an
appropriate orthonormal basis, viz., $\delta {\bf p} = (\delta {p}_{\parallel},
\delta {p}_{\perp}, \delta {p}_{z})$, where $\delta {p}_{\parallel} = \delta
{\bf p}\cdot {\bf p}_0$, $\delta {p}_{\perp}= \delta {\bf p} \cdot ({\hat{\bf
z}} \times {\bf p}_0)$ and $\delta {p}_{z} = \delta {\bf p}\cdot {\hat{\bf z}}$
(similarly for $\delta{\bf g}$). As before we ignore concentration
fluctuations.

Once again, $\delta p_{\parallel}$ is massive, and we rewrite the order
parameter fluctuations in terms of the in-plane splay $\Phi_{\perp} =
\nabla_{\perp} \delta p_{\perp}$ and $\Phi_{z} = \nabla_{z} \delta p_{z}$.
Similarly invoking incompressibility, we rewrite the momentum fluctuations as
the in-plane expansion rate $\Theta_{\perp} =  \nabla_{\perp} \delta g_{\perp}$
and $\Theta_{z} = \nabla_{z} \delta g_{z}$. Eliminating the momentum
fluctuations via force-balance, we find that the resulting linearised dynamical
equations for $\Phi_{\perp}$ and $\Phi_{z}$ give rise to an eigenvalue spectrum
qualitatively resembling (\ref{dispersion}). Thus even in $d=3$, one may define
a stability phase diagram in the $Pe_a - \lambda$ plane; a large enough shear
rate stabilises an orientationally ordered phase.
\\

Note that the vector-ordered phase has a non-zero drift velocity $v_d$, with a magnitude proportional to $c_0 \,p_{0}$ (\ref{steady}). This macroscopic {\it particle current} will result in a counter {\it solvent flow} of the same magnitude. 

Experimental realisations in a planar shear flow geometry will necessarily have to 
contend with finite boundaries; it is therefore important to specify boundary 
conditions for the active order parameter ${\bf p}$.
This is especially important in the polar case, since the local polarisation implies 
a local  
average drift velocity of the active particle relative to the fluid. 
Assuming the active particles cannot penetrate the walls, 
${\bf p}$ must be tangent to the confining walls 
(the homogeneous boundary conditions of liquid crystal physics). 
In plane Couette flow, however, this is at odds with 
the flow-alignment requirement of ${\bf p}$ pointing at 
an angle to the suspension velocity as in Eq. (\ref{steady}). 
This conflict must be resolved by a boundary layer in ${\bf p}$ at 
the walls. The case of wall-normal (homeotropic \cite{degennes}) 
alignment will be discussed elsewhere.

In conclusion we have shown how to stabilise the orientationally ordered phase of 
an active particle suspension by imposing a uniform shear flow. 
We have determined the nonequilibrium phase diagram in the plane of 
``active Peclet number'' and flow-alignment parameter. 
This sets the stage for a study of the unusual rheological features of this 
shear-stabilised oriented phase of active matter \cite{rheo-orient}.
In a forthcoming submission we will use  this framework to study the 
dynamics of reorientation of endothelial cells subject to shear flow
\cite{noria}.  

We thank the Indo-French Centre for a grant (IFCPAR Grant \# 3504-2).

\end{document}